\documentclass[conference]{IEEEtran}
\IEEEoverridecommandlockouts
\usepackage{cite}
\usepackage{amsmath,amssymb,amsfonts}
\usepackage{algorithmic}
\usepackage{graphicx}
\usepackage{textcomp}
\usepackage{xcolor}
\newtheorem{defn}{Definition}
\usepackage{multirow}
\usepackage{graphicx}
\usepackage{balance}

\usepackage{colortbl}
\usepackage{comment}
\definecolor{Gray}{gray}{0.65}
\usepackage[ruled,vlined]{algorithm2e}
\usepackage{subcaption}
\def\BibTeX{{\rm B\kern-.05em{\sc i\kern-.025em b}\kern-.08em
    T\kern-.1667em\lower.7ex\hbox{E}\kern-.125emX}}
    
\begin{document}

\title{DyGCL: Dynamic Graph Contrastive Learning For Event Prediction
}

\author{\IEEEauthorblockN{Muhammed Ifte Khairul Islam}
\IEEEauthorblockA{\textit{Department of Computer Science} \\
\textit{Georgia State University}\\
Atlanta, GA, USA \\
mislam29@student.gsu.edu}
\and
\IEEEauthorblockN{Khaled Mohammed Saifuddin}
\IEEEauthorblockA{\textit{Department of Computer Science} \\
\textit{Georgia State University}\\
Atlanta, GA, USA \\
ksaifuddin1@student.gsu.edu}
\and

\IEEEauthorblockN{Tanvir Hossain}
\IEEEauthorblockA{\textit{Department of Computer Science} \\
\textit{Georgia State University}\\
Atlanta, GA, USA \\
thossain5@student.gsu.edu}
\and
\IEEEauthorblockN{Esra Akbas}
\IEEEauthorblockA{\textit{Department of Computer Science} \\
\textit{Georgia State University}\\
Atlanta,GA, USA \\
eakbas1@gsu.edu}

}

\maketitle
\newcommand{\dypool}{\texttt{DyGCL}}
\begin{abstract}
 Predicting events such as political protests, flu epidemics and criminal activities is crucial to proactively taking necessary measures and implementing required responses to address emerging challenges. Capturing contextual information from textual data for event forecasting poses significant challenges due to the intricate structure of the documents and the evolving nature of events. Recently, dynamic Graph Neural Networks (GNNs) have been introduced to capture the dynamic patterns of input text graphs. However, these models only utilize node-level representation, causing the loss of the global information from graph-level representation. On the other hand, both node-level and graph-level representations are essential for effective event prediction as node-level representation gives insight into the local structure, and the graph-level representation provides an understanding of the global structure of the temporal graph. To address these challenges, in this paper, we propose a Dynamic Graph Contrastive Learning (\dypool) method for event prediction. Our model \dypool\ employs a local view encoder to learn the evolving node representations, which effectively captures the local dynamic structure of input graphs. Additionally, it harnesses a global view encoder to perceive the hierarchical dynamic graph representation of the input graphs. Then we update the graph representations from both encoders using contrastive learning. In the final stage, \dypool\ combines both representations using an attention mechanism, and optimizes its capability to predict future events. Our extensive experiment demonstrates that our proposed method outperforms the baseline methods for event prediction on six real-world datasets.
\end{abstract}


\section{Introduction}
The course of our daily lives is significantly shaped by a variety of human events, including political protests~\cite{ramakrishnan2014beating}, flu epidemics~\cite{deng2020cola, signorini2011use}, and criminal activities~\cite{wang2012automatic}. The proliferation of numerous media outlets has enabled us to apply data-driven machine-learning strategies for predicting future events. This advancement significantly assists local authorities and organizations in making informed decisions and implementing appropriate responses. Recently, deep-learning models have been utilized for event prediction, where features from the input data are autonomously learned. 
Among them, Recurrent Neural Networks (RNNs) specifically treat event data as a time series, learning from the historical context from previous time points and using this knowledge to predict future events. However, RNNs primarily concentrate on the semantic information within the input data without considering the temporal structural dependencies between entities related to the event. To capture the dependency between entities, summarized time-series data are represented as dynamic graphs that have changes in interaction and attributes of nodes over time. While small changes are expected over time, major changes in the structure and attributes of the graph can give important information about the rise of events.

To learn semantic structural information in text by composing graphs, dynamic GNNs have been developed as extensions to standard GNNs to capture the dynamic patterns within dynamic graphs and used for event prediction~\cite{deng2019learning, kosan2021event}. The existing dynamic GNNs for event prediction typically learn the node-level representation of each timestamp's graph based on prior timestamp graphs. More specifically they learn static snapshot node embeddings via a GNNs model that induces the local neighborhood properties to the node representations and their evolution is captured by a recurrent neural network architecture. The final node representations are then employed to predict future events. Although these models show promising results, they primarily focus on node-level representation and overlook the graph-level representation, which is needed to capture the dynamic patterns of input graphs. On the other hand, the node-level representation captures the local structure, while the graph-level representation encapsulates the global structure of the temporal graph. Therefore, for more accurate event prediction, both node-level and graph-level representations are indispensable. 

Recently, self-supervised learning has demonstrated significant success across diverse domains, including natural language processing~\cite{devlin2018bert, brown2020language} and computer vision~\cite{gidaris2018unsupervised, noroozi2016unsupervised}. 
A central concept in this framework, contrastive learning~\cite{chen2020simple}, aims to learn representations that preserve similarity by contrasting two or more semantically coherent views obtained through data augmentations of the same underlying object.
 
A more recent line of research~\cite{lee2022augmentation, hassani2020contrastive,Sun2020InfoGraph: } has explored extending the contrastive framework to graph representation learning and achieved superior performance to regular GNN models. While some models apply different graph augmentation methods, e.g., edge dropping, subsampling, etc., to create different views, some models create different graphs from different types of data, such as an attribute graph in addition to the original graph.

However, despite the high performance of contrastive learning on static graphs, there are only a few methods that use contrastive learning in dynamic graphs for downstream tasks like dynamic link prediction and node classification~\cite{tian2021self, 10184608}. These methods also mainly focused on the local structure of nodes to learn the node representation and ignored the global structure of the dynamic graphs. Existing models use different timestamp graphs as different views of the input graphs as positive and negative samples for contrastive learning. These views mainly represent the local structure of the temporal graphs and are used for learning node representation.

Addressing the aforementioned challenges, this paper introduces a \textbf{Dy}namic \textbf{G}raph \textbf{C}ontrastive \textbf{L}earning (\dypool) method that learns dynamic graph representation for future event prediction. In our model, we have two different dedicated encoders, a local and a global view encoder that learns the local structure and global structure of the input graph, respectively. Our local view encoder captures local neighborhood structural information of the nodes in each time snapshot graph using a Dynamic Graph Convolutional Network via passing node features from the previous time to the current time snapshot graph. In the last time step, it applies a pooling layer to generate one representation for the graph. Furthermore, the global view encoder captures the global structure of the input graphs using dynamic Graph Pooling. We update both graph representations from the two different encoders using contrastive learning, where we maximize the cosine similarity between them. Finally, our model combines both representations using an MLP layer and uses it for future event prediction.  As a result, by combining both encoders, \dypool\ learns a dynamic graph representation that captures local and global dynamic structural patterns of input graphs. To the best of our knowledge, our model is the first of its kind to combine both node-level and graph-level temporal dependency for learning graph representations using contrastive learning for event prediction. 

The main contributions of our work can be outlined as follows:

\begin{itemize}
\item We present an innovative framework called Dynamic Graph Contrastive Learning (\dypool), expressly developed for predicting events. The \dypool\ framework is composed of a local view encoder and a global view encoder. The local view encoder in \dypool\ is responsible for learning the graph representation that captures the local structure of the graph in each time-stamp graph, considering the previous time-stamp representations. In contrast, the global view encoder infers a unified representation for each time-stamp graph and maintains all the previous time-stamp representations to learn the future one. We update both graph representations using contrastive learning, where we maximize the similarity between the two representations.

\item  \dypool\ utilizes both semantic and structural features of the input graph, enhancing the overall effectiveness of the model. Contrary to prior works, which primarily focused on the local structure of input graphs, our approach considers both the local and global structures of input graphs to comprehend the temporal dependency for event forecasting via the local view encoder and the global view encoder, respectively.
\item We compare the results of our model against several deep learning and GNN-based baseline models for event prediction. The experiments demonstrate that our model outperforms these baseline models.
\end{itemize}

The rest of this paper is organized as follows. In section ~\ref{sec:relatedwork}, we discuss the related work on event prediction and how GNN and Dynamic GNN are used for event prediction and Contrastive learning in GNN. In section ~\ref{sec:preli}, we present the preliminary concepts for the problem formulation, Graph Neural Network for Static Graph, and Contrastive Learning. We explain our methodology for \dypool\ and how to incorporate it for event prediction in section ~\ref{sec:method}. In section ~\ref{sec: exp}, we present our results on real-world datasets by comparing current baseline methods. Our final remarks with future work directions are found in section ~\ref{sec:con}.

\section{Related Work}\label{sec:relatedwork}
\subsection{Event prediction}

In the area of event prediction, there is a broad spectrum of real-world applications that have been explored, such as predicting political events~\cite{mueller2018reading,kosan2021event}, forecasting election results~\cite{tumasjan2010predicting}, traffic analysis, trends in the stock market~\cite{bollen2011twitter}, and tracking disease outbreaks~\cite{corley2014disease}. Initial approaches in this field often utilized traditional machine learning techniques. For instance, the use of linear regression~\cite{xiao2011twitter} has been noted for its effectiveness in predicting the timing of future events by analyzing social media data frequency and volume. Furthermore, more complex methods involving paragraph embeddings~\cite{ning2016modeling} and the utilization of topic-specific keywords~\cite{wang2012automatic} have been investigated.

The advancements in event prediction have seen a notable shift from basic machine learning techniques to the adoption of sophisticated deep learning methods, particularly focusing on the temporal aspects of information diffusion. The application of deep neural networks, notably by Ma et al. \cite{ma2015detect, ma2016detecting} using recurrent neural networks, and Liu et al. \cite{liu2018early} combined convolutional and recurrent neural networks, exemplifies this advancement. These models excel in understanding the progression and nuances of events as they unfold, particularly in digital and social media contexts. Xia et al. \cite{xia2020state} further contributed to this field with a model focused on the detailed detection and segmentation of evolving event states, offering a more granular perspective on event dynamics. Despite these technological advancements, a common limitation persists in the predominant focus on semantic information of input data, potentially overlooking other vital aspects, such as contextual and non-semantic factors, which are crucial for a comprehensive and accurate prediction model.

\subsection{Graph-based event prediction}

Graph-based event prediction has seen significant advancements with the implementation of GNNs, which excel in structuring the relationships among words or entities into graphs. The DynamicGCN \cite{deng2019learning} exemplifies this approach by applying a static Graph Convolutional Network (GCN) to input graphs of each time epoch, initializing node features for each epoch with embeddings from the previous one, and starting with word embeddings at the initial time. This model updates node embeddings via a temporal layer, integrating the initial and GCN-derived embeddings, and employs a readout layer to convert these embeddings into a fixed-size vector for event prediction. Similarly,  DyGED\cite{kosan2021event} focuses on learning graph-level representations for each epoch's graph, updated using a recurrent neural network (RNN). This model applies a static GNN and global pooling for learning representations but does not utilize hierarchical graph pooling, which is crucial for capturing the graph's global structure. DyGED then merges all snapshot graph embeddings into a single vector representation for event prediction. While both models mark important contributions to graph-based event prediction, leveraging GNNs to capture complex event dynamics, they also highlight areas for potential improvements, such as the integration of more advanced graph pooling techniques and enhanced processing of temporal information for more robust and accurate predictions.


\subsection{Graph contrastive learning}
Contrastive learning ~\cite{chen2020simple}, a subset of self-supervised learning techniques, has seen widespread adoption in domains like image processing~\cite{gidaris2018unsupervised, noroozi2016unsupervised} and natural language processing. This approach learns data representations by distinguishing between similar (positive) and dissimilar (negative) pairs of data points. The core principle is to align representations of similar samples closely in the latent space while ensuring that dissimilar samples are farther apart. This technique is particularly valuable in scenarios with limited or unlabeled data, as it leverages the inherent structure of the data to learn meaningful representations without relying on external labels.

Graph contrastive learning extends the principles of contrastive learning to the realm of graph representation learning. It addresses the unique challenges posed by graph-structured data, aiming to capture both local node-level and global graph-level structural information. Inspired by Deep InfoMax, Deep Graph Infomax (DGI)~\cite{veličković2018deep} maximizes the mutual information between a node's representation and a high-level summary of the entire graph for learning node representation. This maximization aids in overcoming the limitations of vanilla graph convolutional networks by incorporating a more comprehensive understanding of the graph's global information. Methods like ~\cite{hassani2020contrastive} generate two augmented views of a graph using diffusion matrix and original graph adjacency matrix and learn by bringing representations of the same node in different views closer together. 

DDGCL~\cite{tian2021self} uses contrastive learning for dynamic graphs where it constructs temporal views by sampling nodes from $k$-hop neighborhood at different timestamps. On the other hand, CLDG~\cite{10184608} uses different timespan views of the same graph as a contrastive pair. Both methods mainly focus on the local structure of the graph to learn node representation. Despite challenges like computational intensity and sampling bias, graph contrastive learning has proven effective in unsupervised learning tasks on graphs, offering significant insights for tasks such as node and graph classification and recommendation systems.

Existing Dynamic GCN models or Graph Contrastive models either focus on node-level representation or graph-level representation for event prediction. As a result, they ignore the local or global dynamic structure of the input graphs. In our method, we address this problem and learn node-level and graph-level representation for each time epoch graph. While learning graph-level representations, we apply a hierarchical graph pooling method that helps to capture the global structure of the graph effectively.

\section{Preliminaries and Problem Description} \label{sec:preli}
In this section, we first present some primary concepts and terminology about dynamic graphs and contrastive learning and then define the event prediction problem.

For an event at a specific location, historical event-related articles reveal important information about the rising event. Therefore, these are used to predict future events. In this project, event-related articles are encoded into a sequence of graphs as a dynamic graph where each graph represents the contextual information from a specific timestamp. While nodes in the graph represent words occurring in the articles of that timestamp and edges between nodes represent the occurrence of the words in a predefined fixed-size window. One graph is created for each day and data from $k$ consecutive days is represented as the dynamic graph and is used to learn and predict whether an event will occur on $(k+1)$th day. 

Here, we first define dynamic graphs and then event prediction via dynamic graphs. 

\begin{defn}[\textbf{Dynamic Graph}] A dynamic graph $\mathbb{G}$ is defined as a series of $T$ discrete snapshots denoted as $ \mathbb{G}=\{G_{1}, G_{2}, ..., G_{T}\}$, where $G_{t}$ represents the graph at timestamp $t$. Each $G_t$ has an adjacancy matrix $A_t$ showing the relation between nodes at time $t$
\end{defn}

We define the event prediction problem as the binary classification problem and predict whether there is an event on day $t+1$ using data from t previous days. Formal definition of the event prediction via dynamic graphs is given as follows;
\begin{defn}[\textbf{Event Prediction}] Given a training dataset D, where each sample is represented as a dynamic graph $ \mathbb{G}=\{G_{1}, G_{2}, ..., G_{T}\}$ with initial node feature matrix $H_{t}\in R^{NXd}$ where N is the number of nodes with d dimension at time t, our goal is to learn a graph encoder that maps the input dynamic graph into vector representation and use this representation to predict the future event $\hat{y}$ at time $T+1$.
\end{defn}

For our dynamic graph datasets, which are constructed from the text of event-related articles, we have one global initial node feature matrix $H_{sem}$ representing the semantic meaning of all words appearing over the time obtained with a word embedding model. $H_t$ can be obtained by filtering the words occurring at time $t$.
\\
\subsection{Graph Neural Network for Static Graph}
Recently, GNNs have been used as benchmark models for static graphs for different types of downstream tasks including node classification, link prediction, and graph classification. Graph convolution network (GCN) \cite{Kipf} is the most widely used GNN model. GCN is a multilayer neural network to processes the graph data where it combines the features of each node from its neighbors while propagating the information through the edges. Given an input graph as $G(V, A, H)$ where $V$ is the node-set, $A\in{\mathbb{R^{N\times N}}}$ is the adjacency matrix and $H\in \mathbb{R}^{N\times d}$ is the feature matrix with $d$ dimensional node feature and $N$ is the number of nodes in the graph GCN layer transform the node representation as follows:
\begin{equation}
H^{(1)}=\sigma{(\tilde{D}^{-\frac{1}{2}}\tilde{A}\tilde{D}^{-\frac{1}{2}}H^{(0)}\theta^{(0)})}
\end{equation}
where $\sigma$ is a non linear activation function, $\tilde{A}=A+I$ is the adjacency matrix with self-loop, $\tilde{D}\in{\mathbb{R^{N\times N}}}$ is the normalized degree matrix of $\tilde{A}$, $\theta^{(0)}$ is trainable weight.
GCN model uses multiple convolutional layers to learn the spatial feature for nodes from the connected nodes as follows: 
\begin{equation}
H^{(l+1)}=\sigma{(\tilde{D}^{-\frac{1}{2}}\tilde{A}\tilde{D}^{-\frac{1}{2}}H^{(l)}\theta^{(l)})}
\end{equation}
where $H^{(l)}$ feature matrix and $\theta^{(l)}$ model parameter for $l^{(th)}$ layer. The underlying intuition for each layer is that nodes gather and aggregate information from their local neighbors. With the $l$ layers GCN model, nodes can get information from $l-$hops neighborhood information.
 
 In general, while training the model for the classification problem, cross-entropy loss is used as the supervised loss function, which compares the actual label with the predicted label.\\ 
\subsection{Contrastive Learning}
Contrastive learning (CL) has become one of the most popular approaches for unsupervised representation learning, which learns through comparisons among different samples. This comparison is typically conducted between positive pairs of "similar" inputs and negative pairs of "dissimilar" inputs. Contrastive learning in the graph domain aims to learn node or graph representation for a given input graph by maximizing the similarity between the different views of the input graph in their latent space via contrastive loss. In general, there are three parts to graph contrastive learning: i) View construction, ii) View encoder, and iii) Contrastive loss. 
\begin{itemize}
\item \textbf{View Construction:} A view is represented as graph data, denoted as $vi = (A_{i}, X_{i})$, where $A_{i}\in{\mathbb{R^{N\times N}}}$ and $H_(i)\in \mathbb{R}^{N\times d}$. While it is common to get 2 views, there are also some models that create more than 2 views \cite{chen2023attribute, fang2023gomic}. In practical terms, view augmentation approaches involve techniques like node dropping \cite{zeng2021contrastive}, edge perturbation \cite{zhu2020deep}, attribution masking \cite{you2020graph}, and subgraph sampling \cite{ zhang2020motif}.  
\item \textbf{View Encoder:} View encoder maps the node representation or graph representation from different views of input graphs. As encoders for learning representations from views, any GNN models can be used based on the problem and type of graphs.
\item \textbf{Contrastive loss:} After learning the representations for different views using the view encoder, the contrastive learning model is optimized by a contrastive loss. The objective of the contrastive loss function is to maximize the similarity among different view representations of the same graph and minimize similarity for view representations of other graphs. In general, the contrastive loss can be described as follows:
\begin{equation}
L_{contra} = -log (\frac{sim(Z_{vi},Z_{vj})}{\sum_{j'=1}^{N}sim(Z_{vi},Z_{vj'})})
\label{equ:contraL}
\end{equation}
where the sim(.) function measures the similarity between two representations. We can use any similarity function like cosine similarity and mutual information (MI) in the contrastive loss. $Z_{vi}$ and $Z_{vj}$ are the $i$th view and $j$th view representations of node $v$.

\end{itemize}

\section{Method} \label{sec:method}
\begin{figure*}[ht!]
    \centering{
\includegraphics[width=.89\textwidth]{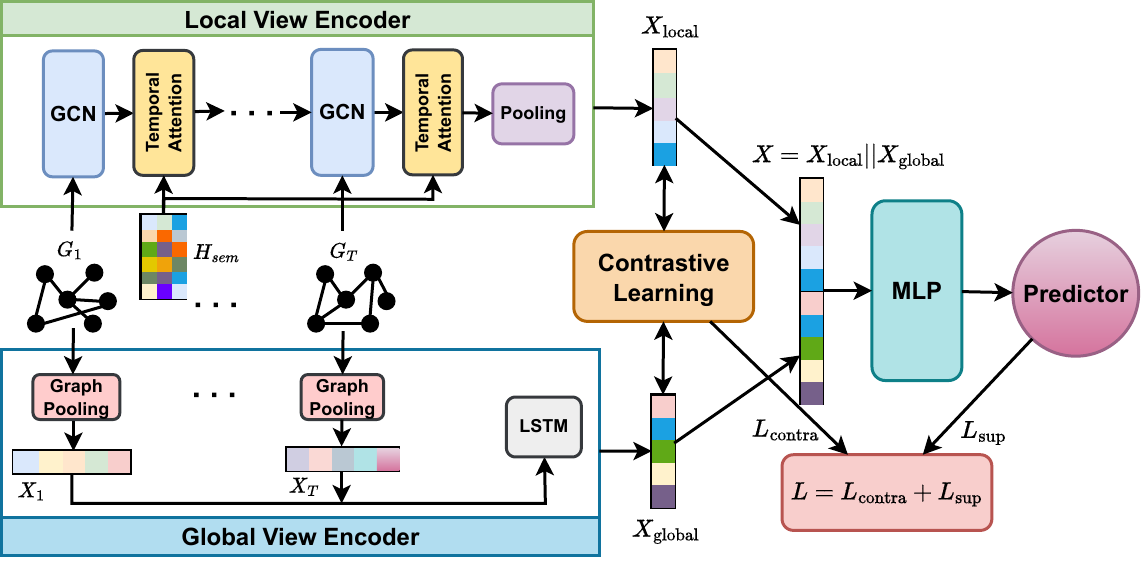}}
    \caption{An overview of Dynamic Graph Contrastive Learning, (\dypool)
 architecture. We feed input graphs into the Local View Encoder to learn the dynamic node representations. In the end, node representations are pooled into a graph-level representation using a pooling layer. We also feed input graphs into the Global View Encoder to learn dynamic graph representation. Use Contrastive learning to maximize the similarity between two graph representations.  Finally, representations from Local View Encoder and Global View Encoder are combined by an MLP layer and feed the representation to the predictor for event prediction. }
    \label{fig:fig1}
\end{figure*}

In this section, we present our proposed Dynamic Graph Contrastive Learning (\dypool) model, including local view and global view encoder with a contrastive loss for event prediction. The overall architecture for the model is presented in Figure~\ref{fig:fig1}. As depicted in Figure~\ref{fig:fig1} and Algorithm~\ref{algorithm:1}, our model is specifically designed to optimize dynamic graph representation learning, particularly for event prediction tasks. The model consists of three main components.

The first component is the local view encoder, where we use a Dynamic Graph Convolutional Network to extract local structural information from the dynamic graphs. Dynamic Graph Convolutional Network extracts dynamic node representations from temporal input graphs. For each temporal graph, it utilizes node embeddings from the preceding time step to enhance the embeddings of the current step. Following this, a temporal attention layer combines the original node embedding with the output of the current step to keep the semantic information in the learning process.

In the second component, we introduce a global view encoder where we use Dynamic Graph Pooling. Dynamic Graph Pooling generates a hierarchical representation of each temporal graph using a hierarchical pooling method. These temporal graph representations are then refined by a recurrent neural network, considering both the current and previous temporal graph representations.

The third component is contrastive learning, which helps update the representations from the two encoders to maximize the similarity between them. Then, these representations are combined and given to an MLP layer for event prediction.
\begin{algorithm}[b!]
\SetAlgoLined
\textbf{Input: } Initial Node Features matrix $H_{(sem)}$, Temporal Graphs, Event Label $Y$\\
\textbf{Output: } Predicted Event label $\hat{y}$\\
${\mathcal{G}} \leftarrow $ randomly sample a batch\\
 \For { $i$ in ${\mathcal{G}}$}
 {
   $A_{T-k,...,T}, H_{sem}= \mathbb{G}_{i}$
   $Z_{local}, H_{L}=LVE(A_{T-k,...,T} ,H_{sem})$\\
    //Local view graph representation\\
   $Z_{global}=GVE(A_{T-k,...,T} ,H_{L})$\\
   //Global view graph representation\\
   
   $L_{\text{contra}} = \frac{Z_{\text{local}}Z_{\text{global}}^{T}}{\|Z_{\text{local}} \| \|Z_{\text{global}}\|}$\\
   //Contrastive Loss\\
   $Z_{\text{local}}=Z_{\text{local}}W_{l}+b_{l}$\\
   $Z_{\text{global}}=Z_{\text{global}}W_{g}+b_{g}$\\
   $Z_{\text{Final}}=MLP([Z_{\text{local}} \| Z_{\text{global}}])$\\
   $\hat{Y}=\sigma(MLP(Z_{Final}))$\\
   $\mathcal{L}_{sup}= -\sum Y log \hat{Y}$\\
   $L= \alpha L_{\text{sup}}+ (1-\alpha) L_{\text{contra}}$
 }

\caption{Dynamic Graph Contrastive Learning (\dypool) }
\label{algorithm:1}
\end{algorithm}
\subsection{Local View Encoder}
 
In general, a graph-level representation is derived from the node-level representations of the graph. Therefore, obtaining an improved graph representation hinges on having superior node representations. To enhance node representation, it is crucial to consider the local structure of nodes, including neighborhood relationships and interactions with neighbors. To learn better node representation by preserving the local structure of the nodes and getting graph-level representation from the node-level representation, we introduce a local view encoder. 

In the local view encoder, we use a Dynamic Graph Convolution Network (DyGCN) that learns dynamic node representation, preserving the temporal local structure for dynamic graphs. It passes the information from the previous time step to the next step to capture the dynamically changing neighboring structures among the input graphs. DyGCN comprises two layers: the static GCN layer and the temporal attention layer. For every time step graph, we apply the static GCN to learn node representations based on the current graph structure and also node features coming from the previous time step. This allows us to encapsulate the local structure of nodes at each time step into vector representations. Within each time step, the static GCN learns the node representation using a message-passing approach, which is described as

\begin{equation}
H^{(t)}=\sigma{(\tilde{D}^{-\frac{1}{2}}\tilde{A}\tilde{D}^{-\frac{1}{2}}H^{(t-1)}\theta^{(t-1)})}
\end{equation}
where $H^{(t)}$ represents the node representation matrix for the $t^{th}$ snapshot graph, $\sigma$ is an activation function, $\tilde{A}=A+I$ is the adjacency matrix with self-loop, $\tilde{D}\in{\mathbb{R^{N\times N}}}$ is the normalized degree matrix of $\tilde{A}$, $\theta^{(t)}$ is the trainable weight for $t^{(th)}$ time step and $H^{(t-1)}$ is the input node representation matrix for ${t}^{th}$ time steps obtained from previous time step's Dynamic GCN. 

\begin{algorithm}[t]
\SetAlgoLined
\textbf{Input: } Initial Node Features matrix $H_{sem}$, Temporal Graphs.\\
\textbf{Output: } Local view Graph Representation $Z_{local}$, Updated node Features $H$\\

  \For {each $t \leftarrow T-k$ \hspace{1.5mm} \text{to} \hspace{1.5mm} $T$}
  {
   $H^{(t)}=\sigma{(\tilde{D}^{-\frac{1}{2}}\tilde{A}^{t}\tilde{D}^{-\frac{1}{2}}H^{(t-1)}\theta^{(t-1)})}$\\
    //Temporal Attention\\
    $H^{(t)'}=H^{(t)}W^{t}_{s}+b^{t}_{s}$\\
    $H_{(sem')}=H_{sem}W^{t}_{0}+b^{t}_{0}$\\
    $H^{(t)}=\tanh([H^{(t)} \| H_{sem}])$\\ 
   }
   $Z_{local}=\frac{1}{N}\sum_{i=1}^{N} [h_{i}]$\\

\caption{Local View Encoder (LVE)}
\label{algorithm:1}
\end{algorithm}

While the topological structure of the graph and its changes over time are important features to detect events, semantic meanings of the words are also important for events. On the other hand, when GCN is applied at each time epoch, it updates node representations with the current time epoch neighborhood. Over multiple timestamps, there is a risk of losing essential semantic details of nodes, potentially leading to the over-smoothing issue. To address this challenge, we create a temporal attention layer. We define initial node features, $H_{sem}$ with their word embeddings, which represent their semantic meanings. 
After obtaining node representations from the current time epochs' GCN, we refine these representations through a temporal attention layer with initial semantic representations, $H_{sem}$. In this layer, we combine the node representation from the current timestamp's GCN and the initial semantic node features using a linear neural network layer. First, we multiply them by learnable weight matrices that signify the importance of each representation for event prediction. Then we concatenate weighted representations and give it to an activation function. Here we use $tanh$ as the activation function. By refining the node representation obtained from GCN with the initial semantic word embedding in each temporal attention layer, we facilitate the fusion of semantic attributes of the nodes. The temporal attention layer applied at each time step is formalized as follows:

\begin{equation}
\begin{aligned}
H^{(t)'}=H^{(t)}W^{t}_{s}+b^{t}_{s}\\
H_{sem'}=H_{sem}W^{t}_{0}+b^{t}_{0}\\
H^{(t)}=tanh([H^{(t)'} || H_{sem'}])
\end{aligned}
\end{equation}
where $H^{(t)}$ is the embedding matrix from GCN layer at time $t$, $H^{(0)}$ is the initial node embeddings or pre-trained word embeddings, $W^{t}_{s}$, $b^{t}_{s}$ and  $W^{t}_{0}$, $b^{t}_{0}$ are learnable perameters for $H^{(t)}$ and $H_{(sem')}$, respectively, and $||$ is the concatenation operation.

After updating node embeddings using the temporal attention layer, we pass it as initial node embedding to the next GCN layer at the time stamp $t+1$.
After getting the final node representation from the Dynamic Graph Convolution layer at the last time step,  we convert the node representations into graph representations using global mean pooling, which takes the average of node features defined as follows:
\begin{equation}
Z_{local}=\frac{1}{N}\sum_{i=1}^{N} [h_{i}]
\label{equ:gpool}
\end{equation}
where $N$ is the number of nodes,  $h_{i}$ is the $i^{th}$ node feature.

\begin{algorithm}[t]
\SetAlgoLined
\textbf{Input:} Node Features matrix $H$, Temporal Graphs.\\
\textbf{Output:} Global view graph representation  $Z_{\text{global}}$\\

  \For{each $t \leftarrow T-k$ \hspace{1.5mm} to \hspace{1.5mm} $T$}
  {
    $S = \sigma(\text{GNN}(H^{t}, A^{t}, \theta_{\text{att}}^{t}))$\\
    $idx = \text{topK}(S, [\alpha \times N])$\\
    $A^{(l+1)} = A_{idx, idx}$\\
    $Z_{t} = \frac{1}{N} \sum_{i=1}^{N} x_{i} \| \max_{i=1}^{N} x_{i}$
   }
   $Z_{\text{global}} = \text{RNN}(Z_{T-k}, \ldots, Z_{T})$\\

\caption{Global View encoder (GVE)}
\label{algorithm:1}
\end{algorithm}

\subsection{Global View Encoder}
The local view encoder is primarily engineered to derive graph-level representation from the dynamic node-level representations from temporal graphs. However, this node-centric approach tends to capture only the local structures of input graphs, often sidelining the global structures.  While local structures show close-by connections, global structures show bigger patterns across the entire graph. For predicting events, these big patterns are also important. Sometimes, events happen because of connections that aren't right next to each other. If we only focus on the local structure, we might miss these important clues. Thus, for predicting events, it is very important to consider global structures. 

To capture higher-order features in input graphs, we introduce a global view encoder where we use a Dynamic Graph Pooling module. At first, we learn a hierarchical graph representation for each time snapshot graph. These graph representations are then refined by a recurrent neural network, considering both the current and previous
temporal graph representations to capture relations between them and also to capture the changes from one to another.

To learn the hierarchical graph representation, selection-based and clustering-based pooling models can be applied. Here, we employ top-$K$ pooling as a selection-based graph pooling method~\cite{pmlr-v97-lee19c}. Selection-based methods are chosen for their memory efficiency and emphasis on the global structure of the input graph. In top-$K$ pooling, we select the top $k$ important nodes from the input graph that are deemed relevant to the event and keep these nodes and relations between them for the next layer. The selection of the top $k$ nodes is determined by calculating attention scores for all nodes, leveraging node features, the GNN model, and an attention parameter, as follows:
\begin{equation}
\begin{aligned}
S=& \sigma(GNN(H^{t},A^{t},\theta_{att}^{t}))\\
 \end{aligned}
 \label{equ:spool}
\end{equation}
Where $S\in \mathbb{R}^{N\times 1}$ represents the node attention scores, $H^{t}$ denotes the node embeddings at time $t$, $A^{t}$ is the adjacency matrix, $\theta_{att}^{t}\in \mathbb{R}^{d\times 1}$ is the learnable parameter matrix at time $t$, and $N$ is the number of nodes.

Following the calculation of attention scores, we select the top $k$ nodes with the highest scores. Subsequently, we construct a new coarse graph using the selected nodes, as outlined below:       
\begin{equation}
\begin{aligned}
idx = & topK(S,[\alpha \times N])\\
A^{(l+1)} = & A_{idx,idx}
 \end{aligned}
 \label{equ:cpool}
\end{equation}
where $idx$ is the indices of top-$k$ nodes, $\alpha$ is the pooling ratio, $N$ is the number of nodes, and $A^{(l+1)}$ is the coarse graph. Multiple graph pooling and GCN layers on each snapshot graph are applied to get a hierarchical graph representation.

As the final layer of the hierarchical graph representation learning, we apply a readout layer to get a fixed-sized graph-level representation for each snapshot graph. The readout function aggregates the node features as follows:
\begin{equation}
Z_{t}=\frac{1}{N}\sum_{i=1}^{N} [h_{i}|| \overset{N}{\underset{i=1}{\max}}\ h_{i}]
\label{equ:four}
\end{equation}
where $N$ is the number of nodes,  $h_{i}$ is the $i^{th}$ node feature and $||$ denotes concatenation. 
The graph pooling method effectively captures the static global structure of the current time epoch graph. Yet, to grasp the dynamic evolution of the global structure in temporal graphs, it is crucial to update the current time epoch graph representation with that of the previous time epoch graphs. For this purpose, a Recurrent Neural Network (RNN) layer is employed. RNNs, well-suited for sequential data, update the current data representation by considering the input data from the preceding steps in the sequence. In our approach, we feed graph representations from the pooling layers into the RNN layer in the following manner:
\begin{equation}
Z_{global}=RNN(Z_{t-k},\dots,Z_{t})
\end{equation}
where $z_{t-k}$ is the graph-level representation of $k$-previous time steps from current time graph. The output of the RNN layer is used as the global graph features.
\subsection{Constrastive Learning}

While the local view encoder and global view encoder extract the different features of the dynamic graph, their final representation should be similar as they belong to the same data. Therefore, we apply contrastive learning to make them similar. Our contrastive learning objective function aims to maximize the cosine similarity or minimize the cosine distance between the graph representations from local and global view encoders. 
In contrastive learning, while one view should be central, we should define positive and negative samples to compare and measure the similarity. 
 We use graph representations from the local view encoder and the global view encoder as positive samples if they belong to the same data. In our model, we do not use any negative sample in contrastive learning as Namkyeong Lee et, al ~\cite{lee2022augmentation} mentioned that contrastive learning on graphs performs better without negative samples. As our objective, we minimize the cosine distance between positive samples. We define our objective function as follows:
\begin{equation}
L_{\text{contra}} = \frac{Z_{\text{local}}Z_{\text{global}}^{T}}{\|Z_{\text{local}} \| \|Z_{\text{global}}\|}
\label{equ:contraLoss}
\end{equation}
Where $Z_{\text{local}}$ is the graph-level representation from local view encoder and $Z_{\text{global}}$ is the graph-level representation from global view encoder

\begin{table*}
\centering
\caption{ \centering{Performance comparisons of our model with baseline models on event prediction } }
\label{table:result}
\normalsize
\begin{tabular}{|c|c |c |c |c  |c|c |c |}
 \hline

 \cellcolor{gray!60}{ \textbf{Method} }
 &\cellcolor{gray!60}{\textbf{Model}}
 & \cellcolor{gray!60}\textbf{Thailand}
 & \cellcolor{gray!60}\textbf{Egypt} & \cellcolor{gray!60}\textbf{Russia} &\cellcolor{gray!60}\textbf{India}& \cellcolor{gray!60}\textbf{NYC Cab}& \cellcolor{gray!60}\textbf{Twitter Weather}
 \\ \hline
 
 & GCN & 76.13 & 75.8 & 76.63  &67.45& 84.91 & 77.21  \\
 
 {Static} & TopKpool &77.03  & 85.28 & 78.45  &65.53& 86.45 & 78.65  \\
 & SAGPool & 77.74 & 86.12 & 80.86  &68.50& 90.82 & 80.78  \\
 & DiffPool & 76.13 & 82.8 & 79.63  &67.48& 88.70 & 76.6  \\\hline
   & GCN+GRU & 79.28 & 83.88 & 79.66  &67.48 & 85.00 & 76.50 \\
 & GCN+LSTM & 78.13 & 83.05 & 79.38 &68.10& 85.07 & 76.55  \\
 & EvolveGCN & - & - & - &-& 84.20 & 78.24  \\

{Dynamic}& DynamicGCN & 80.92 & 84.71 & 84.71  &68.70& 81.00 & 71.30  \\
& DyGED & 73.50 & 85.41 & 81.43  &68.88& 91.20 & 81.00  \\
  & \dypool\ \textbf{(Ours)}& \textbf{86.57} & \textbf{89.28} & \textbf{88.95}  &\textbf{76.85}& \textbf{95.80} & \textbf{90.68}  \\
  
\hline

\end{tabular}

\end{table*}

\subsection{Output layer}
 In the local view encoder, we derive a dynamic node-level representation through the Dynamic GCN module, that predominantly encapsulates the local intricacies of temporal graphs. Following this extraction, we introduce a pooling layer at the culmination of the Dynamic GCN. This layer is responsible for morphing the node embedding into a standardized graph-level embedding, as depicted in Equation ~\ref{equ:four}. Concurrently, from the global view encoder, we utilize a pooling technique to obtain a uniform graph representation, encoding the overarching structure of the graph.
In our approach, the graph-level outputs from both the local view encoder and global are integrated, allowing us to embed both temporal local and global graph structural information into a singular graph-level embedding. The amalgamation of these embeddings is facilitated by an MLP layer. To combine local and global view representations, we first employ two distinct learnable weights for the two embeddings, allocating specialized attention to each. Once the embeddings are individually multiplied by their respective learnable weights, they are concatenated. Subsequently, an activation function is employed to seamlessly merge them, as outlined below

\begin{equation}
\begin{aligned}
Z_{local}=Z_{local}W_{l}+b_{l}\\
Z_{global}=Z_{global}W_{g}+b_{g}\\
Z_{Final}=tanh([Z_{local} || Z_{global}])
\end{aligned}
\end{equation}
where $Z_{local}$ is graph embedding matrix from local view encoder, $Z_{global}$ is graph embedding from global view encoder, $W_{l}$,$b_{l}$, and  $W_{g}$,$b_{g}$ are learnable perameters for $Z_{local}$ and $Z_{global}$, respectively, $||$ is the concatenation operation and $tanh$ is an activation function of the linear layer.

After combining both graph representations into one final graph representation, our model gives it as an input to a multilayer perception layer with the sigmoid function to predict the event occurrence and calculate the supervised loss as follows:

\begin{equation}\begin{aligned}\label{equ:lossf}
\hat{Y}=\sigma(MLP(Z_{Final}))\\
\mathcal{L}_{sup}= -\sum Y log \hat{Y}
\end{aligned}
\end{equation} 
where  $Z_{Final}$ is the graph representation, $\hat{Y}$ is the predicted event occurrence, and $Y$ is the actual event occurrence.

We jointly train our model with a weighted sum of the supervised loss and contrastive loss as follows:
\begin{equation}
L= \alpha L_{\text{sup}}+ (1-\alpha) L_{\text{contra}}
\label{equ:eleven}
\end{equation}
 where $\alpha$ is a hyperparameter of the model.

\begin{table}
\normalsize
\centering
\caption{Dataset statistics. $|S|$ is the number of samples, $\bar{N}$ and $\bar{E}$ are the average number of nodes and edges, respectively, and $|Ev|$ is the number of events.}
\label{table:table1}
\begin{tabular}{c c c c c  }
 \hline 	
\cellcolor{gray!60}\textbf{Datasets} &\cellcolor{gray!60}$|S|$&\cellcolor{gray!60}$\bar{N}$ 
 &\cellcolor{gray!60} $\bar{E}$ &\cellcolor{gray!60} $|Ev|$ \\\hline\hline
Thailand &1883 &600 & 7281&715\\\hline

Egypt & 3788 & 675 &9680 & 1469  \\\hline
Russia & 3552 & 645 &9776 & 1171 \\\hline
India & 12249 & 685 &12994 & 4586 \\\hline
NYC Cab & 4464& 263 &3717 & 162 \\\hline
Twitter & 2557& 1000 &10312 & 287  \\\hline
\end{tabular}

\end{table}

\section{Experiment} \label{sec: exp}
In assessing our model's performance for the event prediction task, treating it as a dynamic graph classification problem, we conduct a comprehensive evaluation. We compare our model's performance against six distinct baseline models. Additionally,we delve into the analysis of the impact of the number of historical days on event prediction. Furthermore,  we present results for variations of our model, incorporating different message-passing models and graph-pooling methods. We also visualize the global structure of temporal graphs.

\subsection{Datasets} In our experiments, we use six datasets. Among them, four of them are social event datasets, one is a weather event dataset and the other is a traffic event dataset. Table ~\ref{table:table1} shows the statistics of six datasets. 

\textbf{Thailand}, \textbf{Egypt}, \textbf{Russia}, and \textbf{India} event datasets are collected from the Integrated Conflict Early Warning System (ICEWS)\cite{https://doi.org/10.7910/dvn/28075, deng2019learning}. These datasets include information on political events and are designed for assessing both national and international crises. We concentrate on data sourced from major cities such as Delhi, Mumbai, Kolkata, and others in India, Bangkok in Thailand, Cairo in Egypt, and Moscow in Russia. Rallies, strikes, violent protests, and passage obstructions are frequent event types on these datasets.

The \textbf{NYC Cab} represents a mobility network that contains geo-tagged mass-gathering events, such as concerts and protests. The \textbf{Twitter Weather} dataset showcases user-generated content pertaining to weather events, including storms and earthquakes.

\begin{figure*}
    \centering
    \begin{subfigure}[h]{0.45\textwidth}
        \centering
        \includegraphics[width=0.9\textwidth]{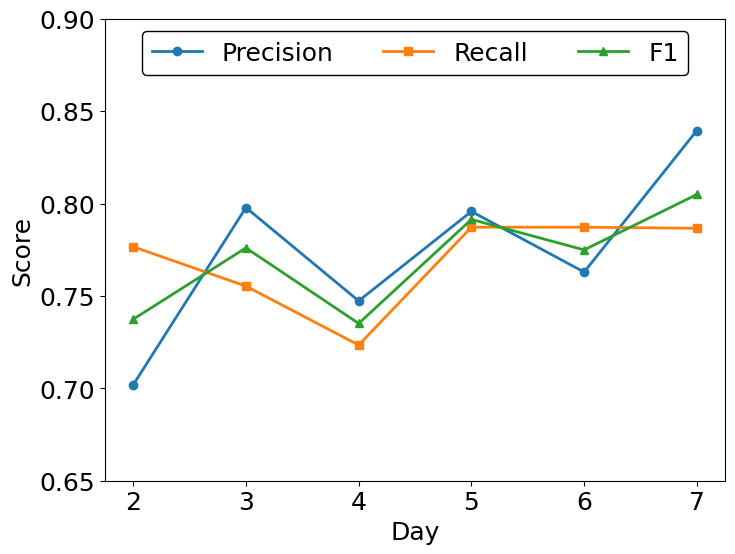}
        \caption{Thailand}
    \end{subfigure}%
    ~     
    \begin{subfigure}[h]{0.45\textwidth}
        \centering
        \includegraphics[width=0.9\textwidth]{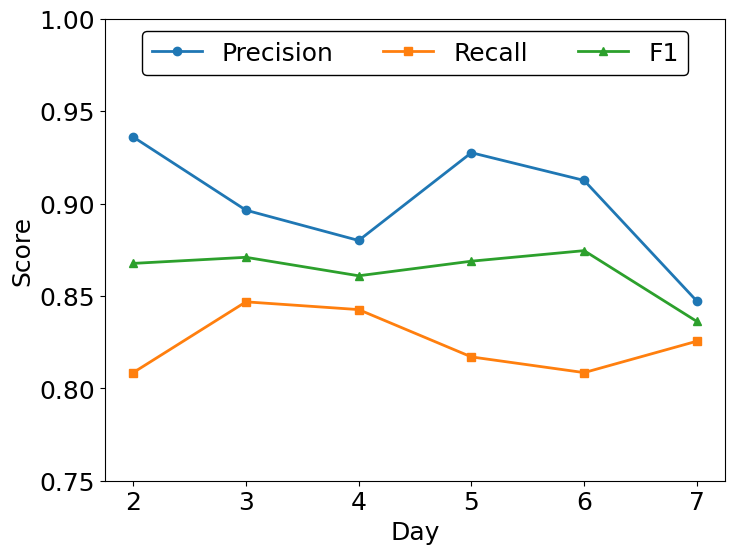}
        \caption{Egypt}
    \end{subfigure}%
    ~

    \caption{Detailed Event prediction results of $\dypool_{sup}$\ model without contrastive learning for the different number of historic days.}
    \label{fig:history}
\end{figure*}

\begin{figure*}
    \centering
    \begin{subfigure}[h]{0.45\textwidth}
        \centering
        \includegraphics[width=0.9\textwidth]{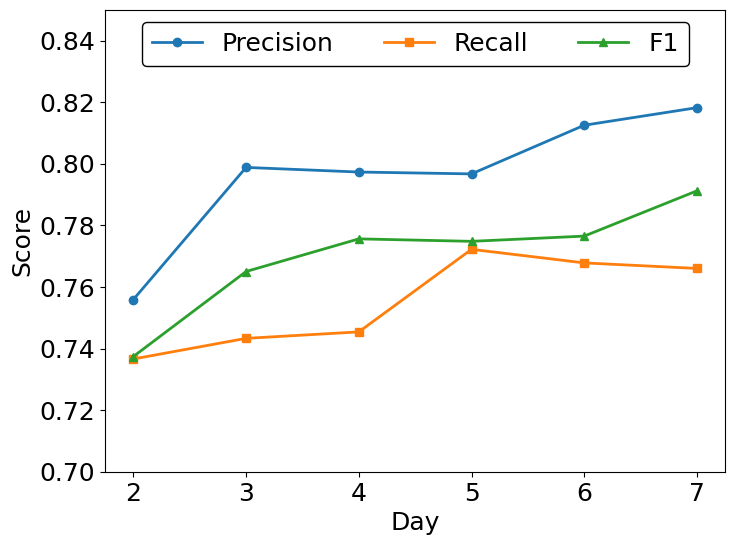}
        \caption{Thailand}
    \end{subfigure}%
    ~     
    \begin{subfigure}[h]{0.45\textwidth}
        \centering
        \includegraphics[width=0.9\textwidth]{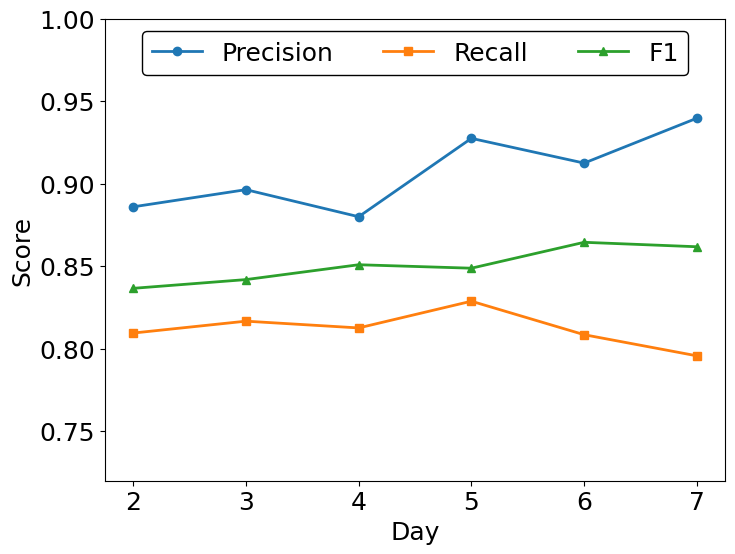}
        \caption{Egypt}
    \end{subfigure}%
    ~

    \caption{Detailed Event prediction results of \dypool\ model for the different number of historic days.}
    \label{fig:historyContra}
\end{figure*}

\subsection{Baseline}
We use several graph neural network methods as baseline methods which mainly focus on Dynamic graphs or static graph classification. We divide our baseline methods into two classes: (1) Static methods, and (2) Dynamic methods. 

\subsubsection{Static Methods}
We compare our model with GCN\cite{Kipf}, TopKpool\cite{gao2019graph},  SAGPool\cite{pmlr-v97-lee19c} and DiffPool \cite{ying2018hierarchical}, which are most common graph representation learning methods. For the static methods, we combine all time step graphs in a dynamic graph into one graph for each sample. 

\begin{itemize}
\item \textit{GCN}:  Graph Convolutional network is the most popular GNN model for node representation learning. GCN aggregates the neighborhood information to update the node representation. After getting node representation we apply global pooling described in equation~\ref{equ:gpool} to get graph-level representation. 
\item \textit{TopKPool}: TopKPool is a hierarchical graph pooling method that selects top-$k$ nodes for pooling operation. It considers the topological structure of the graph to select the top-$k$ nodes.
\item \textit{SAGPool}: SAGPool is also a hierarchical graph pooling method that selects top-$k$ nodes using the self-attention method to select top-$k$ nodes for pooling operation. 
\item \textit{DiffPool}: DiffPool is a cluster-based hierarchical graph pooling method. It uses a Graph Neural Network to learn a cluster assignment matrix of the input graph. Then it uses the cluster assignment matrix for graph pooling where it combines the nodes in each cluster into supernodes for the next layer.
\end{itemize}

\subsubsection{Dynamic Methods}
 For the dynamic baseline models, we use GCN+LSTM~\cite{GCNGRU}, GCN+GRU~\cite{GCNGRU}, DynamicGCN~\cite{deng2019learning}, EvolveGCN~\cite{pareja2020evolvegcn} and DyGED~\cite{kosan2021event}.  

 \begin{itemize}
     \item \textit{GCN+LSTM}: This model learns the dynamic graph representation using GCN and LSTM models. GCN+GRU is a variation of this model where it replaces LSTM with the GRU model. It is designed as a temporal graph neural network to predict traffic conditions in the traffic network. 
     \item \textit{DynamicGCN}: This model applies GCN and a temporal layer for each snapshot to learn dynamic node representations. It converts dynamic node representation to graph representation using a mask linear layer. It is designed for event prediction on social media data.
     \item \textit{EvolveGCN}: EvolveGCN is a dynamic graph convolution network that uses GCN and Recurrent Neural Network (RNN) to learn the representation of dynamic networks. It transfers the parameter matrix from one timestamp's GCN to another timestamp's GCN. It is a more general model that is applied to edge prediction, edge classification, and node classification for dynamic graphs. Here we also use the global pooling method in the last layer of the model to convert node representation to graph representation.
     \item \textit{DyGED}: DyGED is the most recent event prediction model that uses global graph pooling on each time step graph and applies RNN models to include macro-level graph dynamics for graph representation learning. 
 \end{itemize}

\begin{figure*}\begin{center}
    
    \begin{subfigure}[h]{0.45\textwidth}
        \centering        \includegraphics[width=0.9\textwidth]{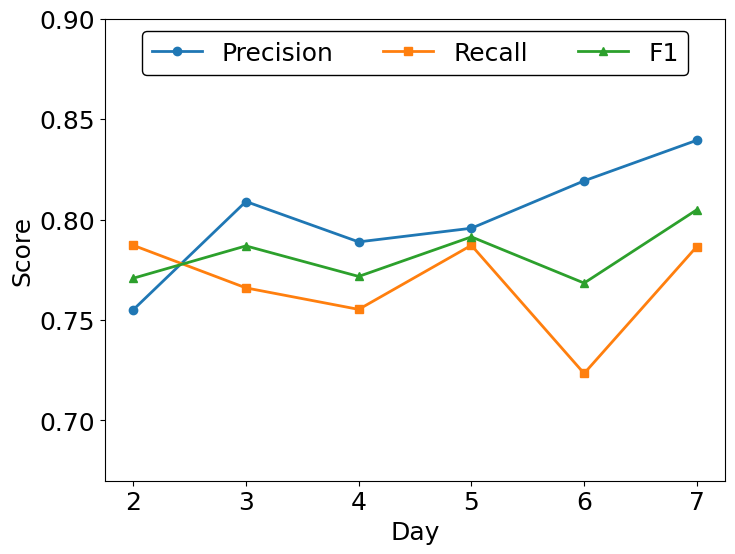}
        \caption{Thailand}
    \end{subfigure}
~
        \begin{subfigure}[h]{0.45\textwidth}
\centering\includegraphics[width=0.9\textwidth]{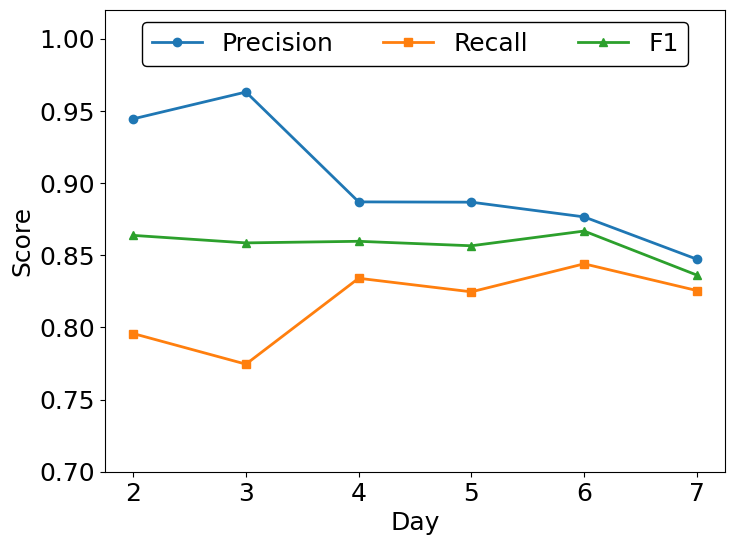}
        \caption{Egypt}
    \end{subfigure}%
\end{center}

    \caption{Detailed event prediction results of $\dypool_{sup}$\ model without contrastive loss for the different number of lead days.}
    \label{fig:lead}
\end{figure*}

\begin{figure*}
       \begin{center}
           
    \begin{subfigure}[h]{0.45\textwidth}
        \centering
\includegraphics[width=0.9\textwidth]{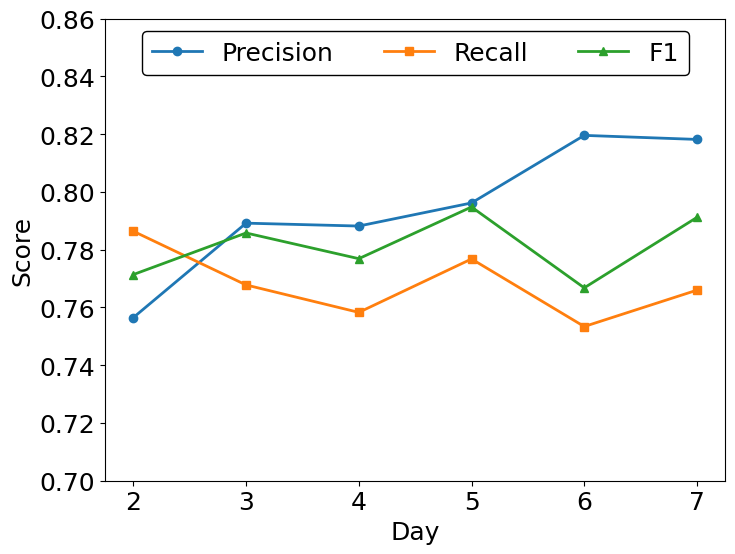}
        \caption{Thailand}
    \end{subfigure}
~
        \begin{subfigure}[h]{0.45\textwidth}
        \centering
        \includegraphics[width=0.9\textwidth]{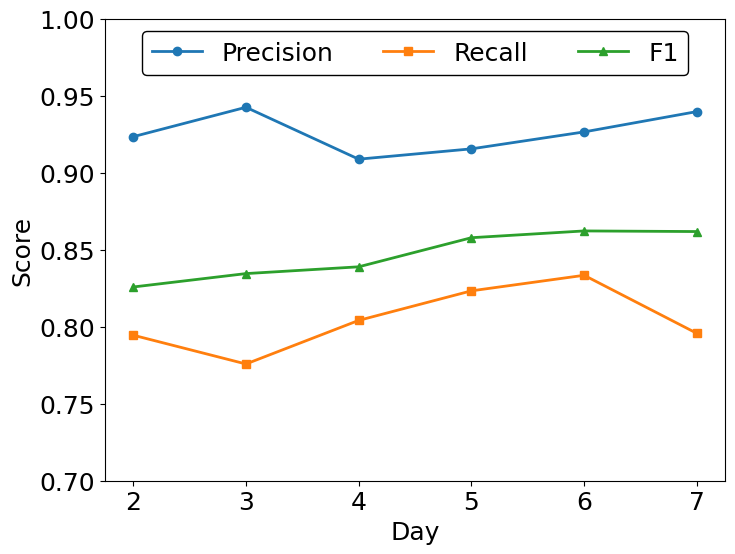}
        \caption{Egypt}
    \end{subfigure}%
   
       \end{center}

    \caption{Detailed event prediction results of the \dypool\ model for the different number of lead days.}
    \label{fig:leadContra}
\end{figure*}

 \subsection{Experimental Settings} To assess our model for the event prediction task, we partition the data into three segments: 70\% for training, 15\% for validation, and 15\% for testing. This splitting process is repeated ten times using ten different random seeds. We take the average of 10 different runs as the final results. Our model is implemented using PyTorch and PyTorch Geometry library, with the optimization performed by the Adam Optimizer. In order to determine optimal hyperparameters, we conduct grid search within the specified ranges: learning rate in $\{1e-2, 5e-2, 1e-3, 5e-3, 1e-4, 5e-4\}$, weight decay in $\{1e-2, 1e-3, 1e-4, 1e-5\}$, pooling ratio in $\{1/2, 1/4\}$, dropout ratio fixed at 0.2, and hidden size in $\{16, 32, 64, 128\}$.

The training process halts if the validation loss fails to improve for 50 epochs. For the initial node embeddings of the dynamic graph, we use a 100-dimensional word embedding vector obtained through the Word2Vec method.

 \subsection{Result} In Table~\ref{table:result}, we present a summary of our event prediction accuracy results, with results of  Static and Dynamic baseline models. All models were run ten times with ten random seeds, and average accuracy was calculated for event prediction. We treat the event prediction task as a binary graph classification problem, where class labels are $\{0,1\}$ indicating whether an event occurs on that day or not. The EvolveGCN model is applicable to datasets where all samples have the same number of nodes. This constraint limits the availability of results to only NYC Cab and Twitter weather datasets that meet this requirement. For other datasets, the number of nodes is different for different samples. 
 
 As we can see from Table~\ref{table:result}, our model consistently works well and outperforms all baseline models across all datasets. Notably, our model improves the highest accuracy of baselines by 6.9\% for the Thailand dataset with 86.57\% accuracy, 3.1\% for Egypt with 89.95\% accuracy, 4.2\% for Russia with 88.96\% accuracy, 11.5\% for India datasets with 76.85 accuracy, 4.6\% for NYC Cap dataset with 95.80 accuracy and 9.6\% for Twitter Weather dataset with 90.68 accuracy. For political datasets (Thailand, Egypt, Russia, India), the DynamicGCN model achieves the second-highest accuracy, except for the Egypt dataset where SAGPool yields the second-highest accuracy. For NYC Cab and Twitter weather datasets, DyGED gets the second-highest accuracy. 

Additionally, we observe that dynamic methods generally outperform static baseline methods for all datasets, with the exception of the Egypt dataset where SAGPool, as the static graph pooling method, performs better than other baselines. This result suggests the importance of global structural information for the Egypt dataset. Importantly, the graph pooling method consistently demonstrates superior accuracy for these two datasets, indicating the significance of higher-order structural information.

\textbf{Sensitivity Analysis:} In our previous experiment, we used the previous 7 days' data to make a prediction on Day 8. In this experiment, we investigate the prediction performance by varying both the number of prior days and lead time. To see the effect of contrastive learning on prediction performance for different numbers of days time we run our original model \dypool and $\dypool_{sup}$ where we directly combine the representations of local and global view encoders using the MLP layer. We remove the contrastive loss $L_{contra}$ from the model and use only the supervised loss $L_{sup}$ to optimize the model.

 Historical days denote the previous number of days used as the training data. We want to see whether there is an effect of the number of historical days preceding the event on the event prediction task by changing it from 2 to 7. In Figure~\ref{fig:history} and Figure~\ref{fig:historyContra}, we present precision, recall, and F1-score for different numbers of days for the event using \dypool\ and $\dypool_{sup}$ models. In these figures, ``day2" denotes the use of data from the last two preceding days ($6^{th}$ and $7^{th}$ days) for predicting events on the $8^{th}$ day. Similarly, for ``day3", we use $5^{th}$, $6^{th}$, and $7^{th}$ days, and for other days.
Contrary to the assumption that training with more days always leads to better results, figure ~\ref{fig:history} shows that this is not always true. Figure~\ref{fig:history} shows the result for the $\dypool_{sup}$ model on the Thailand and Egypt datasets. From the figure, we can see that for the Thailand dataset, we get better results when we include all 7 days of information. We can also observe that using fewer days also gives more accuracy than a higher number of days. For example, for ``day 3" where we use 3 days of information for event prediction, we get more accuracy than ``day 4" where we use 4 days of information for event prediction.  In contrast, better results are obtained with a fewer number of historic days in the case of the Egypt dataset. We get the highest score for ``day 2" while we get the lowest score for ``day 7". 

Figure~\ref{fig:history} shows the result for the \dypool\ model for the Thailand and Egypt datasets. For this model, we can see that for both datasets the Precision and F1-score increase gradually increase when we increase the number of historic days for event prediction. For both datasets, we get the highest score for ``day 7". 
\begin{table}[h]
\centering
\caption{ \centering{Performance comparisons of different GNN models on \dypool\ for Thailand and Egypt datasets.} }
\normalsize
\begin{tabular}{|c|c| c c c |} 
 \hline
 
{Dataset}&
 {Method}
 &   \cellcolor{gray!25}\textbf{Precision} & \cellcolor{gray!25}\textbf{Recall} & \cellcolor{gray!25}\textbf{F1}

 \\ \hline
 
  \multirow{3}{*} {Thailand}& GCN  &  \textbf{0.8182} & 0.766 & \textbf{0.7912}   \\
 & GAT & 0.75 & \textbf{0.798} & 0.773  \\
  &GraphSAGE& 0.779 & 0.787 & 0.783  \\

\hline

  \multirow{3}{*} {Egypt}& GCN  &   \textbf{0.9397} & 0.7957 & \textbf{0.8618}  \\
  &GAT & 0.834 & \textbf{0.855} & 0.814  \\
  &GraphSAGE& 0.864 & 0.835 & 0.847  \\

\hline
\end{tabular}
\label{GNN} 
\end{table}

The lead time indicates the number of days in advance a model makes predictions of an event. In our experiment, we also explore the effect of the lead day on the model performance by varying the number of lead days from 1 to 7.  We present precision, recall, and F1-score for different numbers of days for the event using \dypool\ and $\dypool_{sup}$ models in Figure~\ref{fig:lead} and Figure~\ref{fig:leadContra}. The lead day signifies how many days in advance our model predicts the event. As an example, when we refer to "Day6," it implies the utilization of data from the $1^{st}$ day to the $6^{th}$ day for predicting the event on the $8^{th}$ day, indicating a two-day advance prediction. In both datasets, we observe that the optimal lead time varies for different scoring metrics. From figure~\ref{fig:lead}, we can see that the $\dypool_{sup}$ model provides the highest scores for Precision, Recall, and F1-score with ``day 7" for the Thailand dataset. For the Egypt dataset, we get the highest Precision score for ``day 2"  and for Recall and F1-score we get the highest score for ``day 6". We can see that for all three scores, our model gives the lowest score for ``day 7".  Notably, for all three scores, our model shows the lowest score for ``day 7," suggesting that adding more days' information may introduce irrelevant details to the event.

In Figure~\ref{fig:leadContra}, the performance of the \dypool\ model with contrastive loss is illustrated for different numbers of lead days. Notably, for Precision and F1-score, \dypool\ with ``day 7" yields the highest scores for both datasets. In this scenario, \dypool\ performs better with a higher number of lead days' information compared to fewer days. For the Thailand dataset, we get the highest Recall score for ``day 5" and for The Egypt dataset \dypool\ gives the highest Recall score for "day 6".    

\begin{table}[h]
\centering
\caption{ \centering{Performance comparisons of different graph pooling models on \dypool\ for Thailand and Egypt datasets.} }
\normalsize
\begin{tabular}{|c|c| c c c | } 
 \hline
 
 Dataset&
 {Method}

  & \cellcolor{gray!25}\textbf{Precision} & \cellcolor{gray!25}\textbf{Recall} & \cellcolor{gray!25}\textbf{F1}

 \\ \hline
 
 \multirow{3}{*} {Egypt}& Top-$K$  & \textbf{0.839} & \textbf{0.787} & \textbf{0.805}  \\
  &SAGPool& 0.8182 & 0.766 & 0.7912  \\
 & DiffPool & 0.8161 & 0.7553 & 0.7845  \\

\hline
\multirow{3}{*} {Egypt}& Top-$K$   & 0.847 & 0.826 & 0.836  \\
  &SAGPool& \textbf{0.9397} & 0.7957 & \textbf{0.8618}  \\
 & DiffPool & 0.8664 & \textbf{0.8553} & 0.8608  \\

\hline
\end{tabular}
\label{pooling} 
\end{table}

\textbf{Ablation Study:} In our ablation study, we investigate the impact of different parts of the model including contrastive learning, the GNN model in the local view, the RNN layer, and the pooling layer in the global view, on the model performance. We present results for all ablation studies for Thailand and Egypt datasets as the representative except contrastive learning where we present results for all datasets.

We first experiment with GNN models. In addition to GCN as the default model, We apply two other popular GNN models, which are GAT and GraphSage to see the effect of them on the model performance. Table~\ref{GNN} shows the Precision, Recall, and F1 scores for different GNN models we use in our model to learn node representation for each snapshot graph in the local view. From the table, we can see that the GNN model to learn the node representation has a big impact on the model performance. For both datasets, the GCN model performs much better than the other models, especially with respect to precision. Also, the GraphSAGE model gives better results than the GAT model.

\begin{figure*}[t]\begin{center}
    
    \begin{subfigure}[h]{0.235\textwidth}
        \centering        \includegraphics[width=1.0\textwidth]{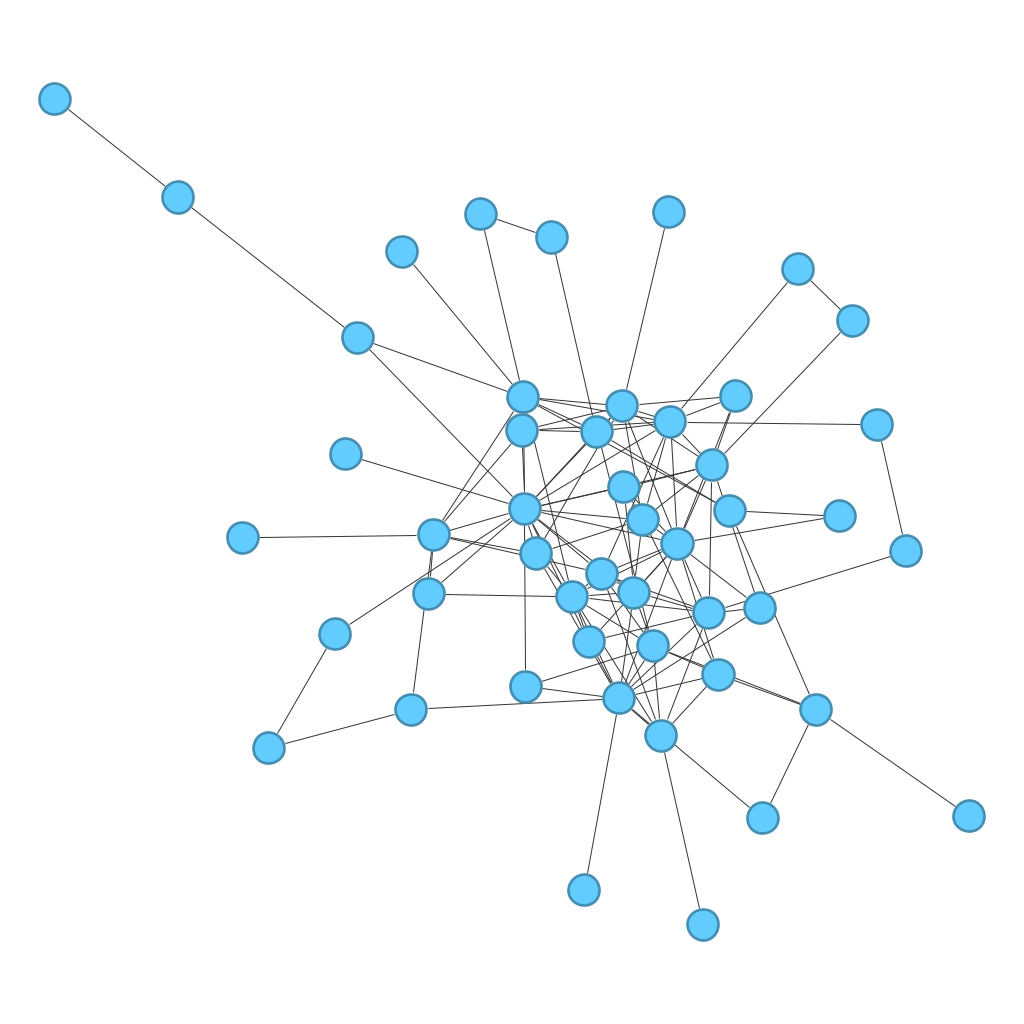}
        \caption{T-4}
    \end{subfigure}
~    
    \begin{subfigure}[h]{0.235\textwidth}
        \centering        \includegraphics[width=1.0\textwidth]{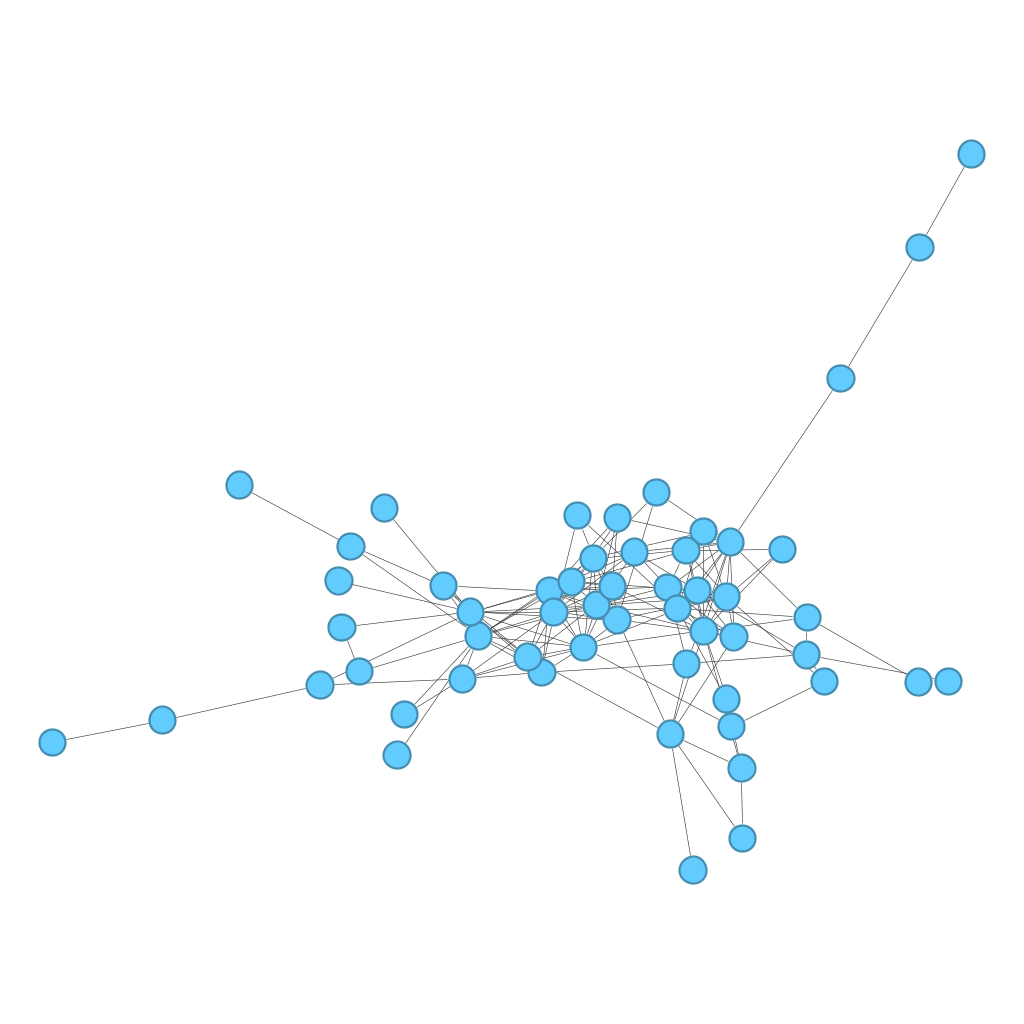}
        \caption{T-3}
    \end{subfigure}
~
        \begin{subfigure}[h]{0.235\textwidth}
\centering\includegraphics[width=1.0\textwidth]{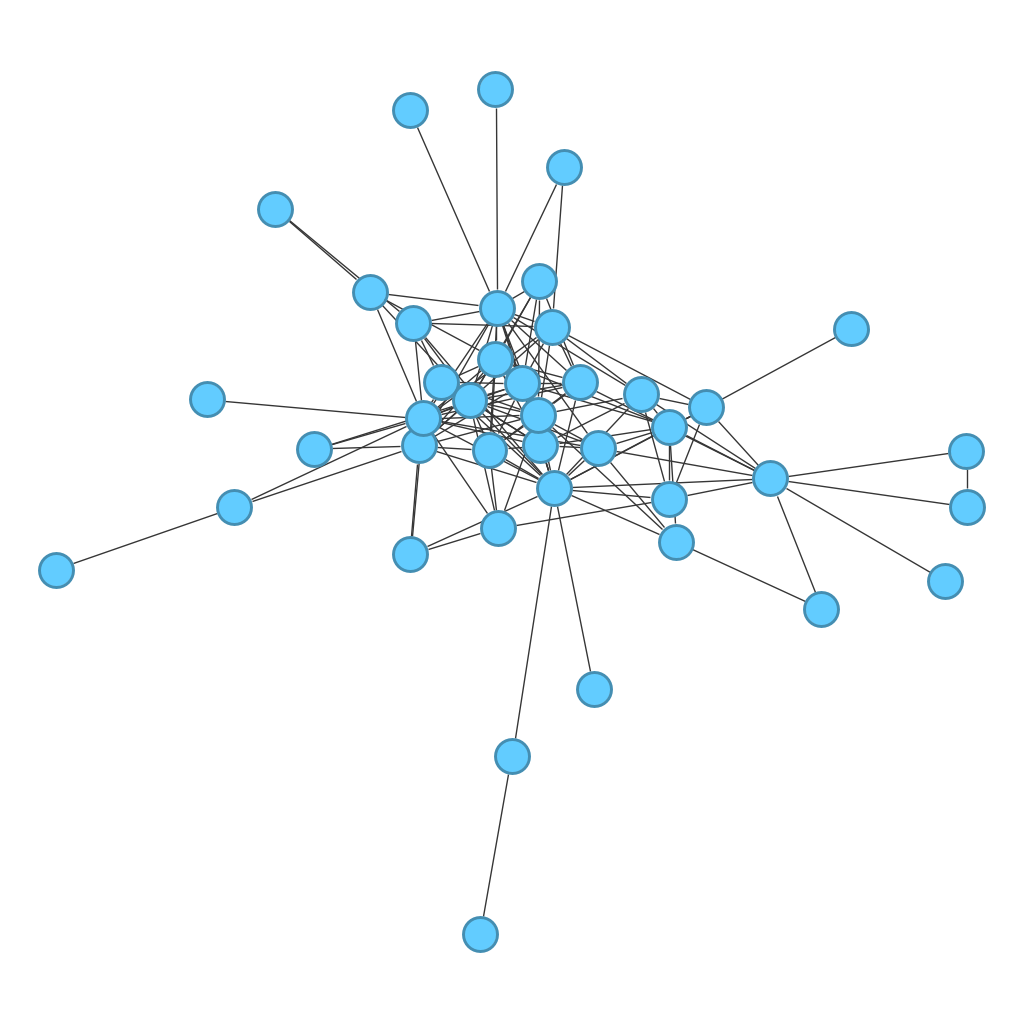}
        \caption{T-2}
    \end{subfigure}
~
        \begin{subfigure}[h]{0.235\textwidth}
\centering\includegraphics[width=1.0\textwidth]{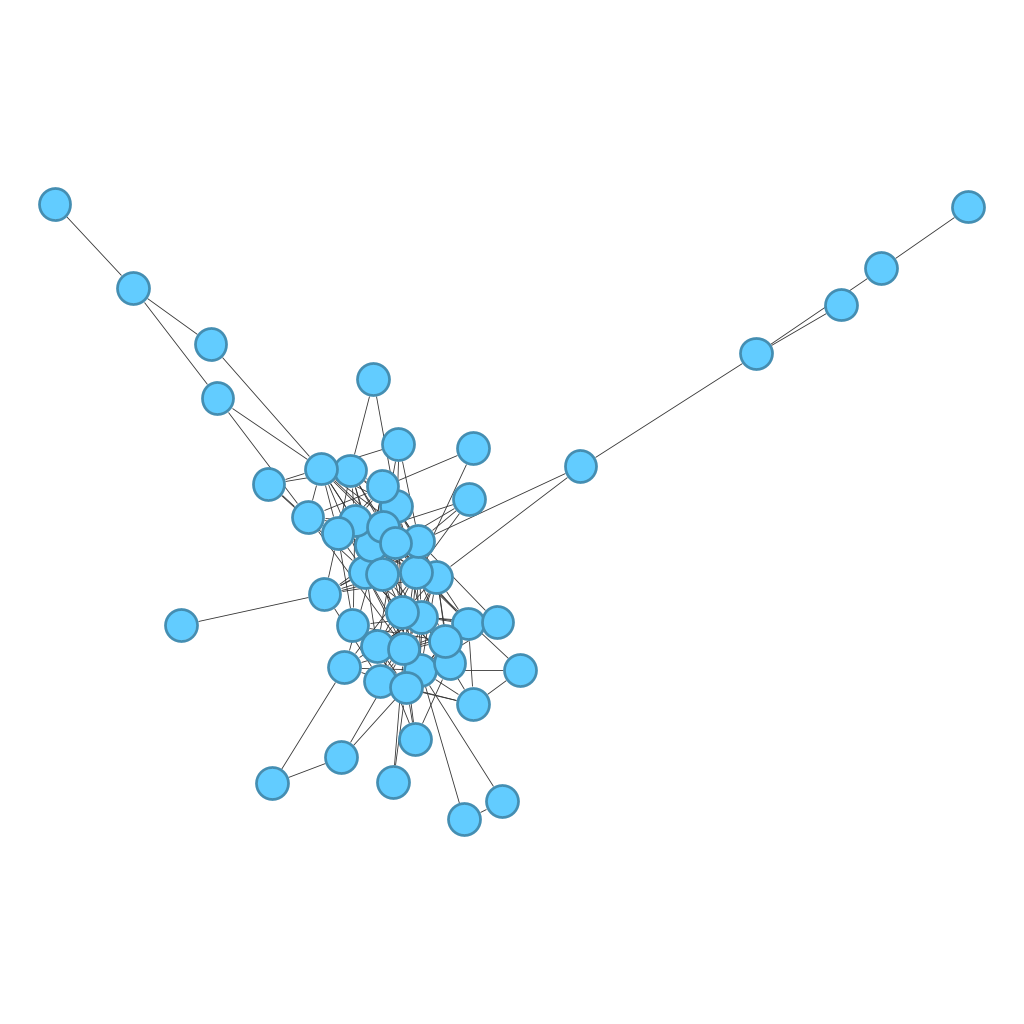}
        \caption{T-1}
    \end{subfigure}
   
\end{center}
    \caption{Temporal pooled graphs in Global View Encoder for a sample event in NYC cab dataset. }
    \label{fig:vis}
\end{figure*}

We also change the hierarchical graph pooling method in the global view encoder to see the effect of different graph pooling methods. In addition to Top-$K$ pooling as the default one, we select SAGPool as the selection-based graph pooling method and DiffPool as the cluster-based pooling method. Table~\ref{pooling} shows the Precision, Recall, and F1-score for three different graph pooling methods for the Thailand and Egypt datasets. From the table, we can see that the pooling method has a big impact on the model performance. In general, selection-based methods outperform the cluster-based method. For the Thailand dataset, the Top-$k$ graph pooling method gives the highest accuracy and for the Egypt dataset, the SAGPool performs better than other methods. 

\begin{table}[t]
\centering
\caption{ \centering{Performance comparisons of different RNN models on \dypool\ for Thailand and Egypt datasets.} }
\normalsize
\begin{tabular}{|c|c| c c c |} 
 \hline

 {Datasets}
 &
 {Method}

  & \cellcolor{gray!25}\textbf{Precision} & \cellcolor{gray!25}\textbf{Recall} & \cellcolor{gray!25}\textbf{F1} 
 
 \\ \hline
 
   \multirow{3}{*} {Thailand}&LSTM  &  0.8182 & \textbf{0.766} & \textbf{0.7912}  \\
  &GRU & \textbf{0.8514} & 0.6702 & 0.75  \\
 & Transformer& 0.7765 & 0.7021 & 0.7374  \\ \hline
 \multirow{3}{*} {Egypt}&LSTM  &\textbf{0.9397} & 0.7957 & 0.8618 \\
 &GRU  & 0.9151 & 0.8255 & 0.8680 \\
 &Transformer  &  0.9112 & \textbf{0.8298} & \textbf{0.8686} \\
\hline

\end{tabular}
\label{RNN} 
\end{table}

In addition, we use different Recurrent Neural Networks(RNN) in the global view encoder to see their effect on them for the event prediction task. While LSTM is the original RNN model in the global view, we use two different popular RNN models, which are GRU, and Transformer in the global view encoder. Table~\ref{RNN} shows the Precision, Recall, and F1 scores for different RNN models. As we can see from th table, RNN model has an impact on the model performance, especially for the Thailand dataset. For the Thailand dataset, LSTM gives the highest Recall and F1 scores with around 0.6 increase.   GRU gives the highest Precision value. On the other hand, the Transformer model performs better than the LSTM and GRU models for the Egypt dataset where it gives the highest Recall and F1 score. However, the difference is not high enough to say that it is the best model.

\begin{table}[t]
\centering
\caption{ \centering{Performance comparisons with contrastive loss (\dypool) and without contrastive loss ($\dypool_{sup}$).} }
\label{table:result2}
\normalsize
\begin{tabular}{|c|c |c |}
 \hline
 
\cellcolor{gray!60} \textbf{Dataset} & \cellcolor{gray!60}$\textbf{\dypool}$ & \cellcolor{gray!60} $\textbf{\dypool}_{sup}$ \\\hline
 
 Thailand & \textbf{86.57} & 85.87   \\
 
  Egypt & \textbf{89.28}  & 86.64   \\
  Russia & \textbf{88.95} & 88.37   \\
  India & \textbf{76.71} & 76.85   \\
  NYC Cab & \textbf{90.68} & 89.88  \\
  Twitter weather & \textbf{95.80} & 95.74  \\\hline

\end{tabular}

\end{table}

Furthermore, we investigate the impact of contrastive loss on event prediction. We run \dypool\ and $\dypool_{sup}$\ for all datasets and report the result in Table~\ref{table:result2}. Different than other ablation studies, we present results for all datasets. As we can see from the table, the model with contrastive loss \dypool\ gives better accuracy than the model without contrastive loss $\dypool_{sup}$. For Thailand, Egypt, NYC Cab, and Twitter weather dataset \dypool\ model gives around 1\%, 3\%, 1\%, 1\%, and 3\% more accuracy respectively than the $\dypool_{sup}$ model. For the Russia and India datasets, the accuracy of both models is very close where the difference is less than 1\%.

\textbf{Visualization:} We also visualize the temporal graphs to see the global structure of the temporal graph before an event. For this experiment, we select a sample from the NYC Cab dataset where an event has occurred. We took four days' temporal graph before that event and applied our pre-trained  \dypool model. Then we get the pooled graph from the output of hierarchical graph pooling layers in the global view encoder and present the graphs in Figure~\ref{fig:vis}. As we can see from the figure, the global structure of temporal graphs before the event is changing a lot and our model is capturing those structures with the local and global view encoders.

\section{Conclusion and Future Work }\label{sec:con}
In this work, we propose a Dynamic Graph Contrastive Learning model \dypool\ for event prediction. There are two view encoders in our model one is a local view encoder and another one is a global view encoder. The local view encoder learns dynamic graph representation by capturing the temporal local structure of input graphs and the global view encoder learns dynamic graph representation that encodes the temporal global structure of the input graph. We update both representations using contrastive learning where the objective is to maximize the similarity between two representations. Then our model combines the representations from the local view encoder and the global view encoder into a fixed-sized vector representation to capture dynamic local and global patterns among the temporal graphs. Finally, our model predicts events using this fixed-sized vector representation. In the experiment, we show that our model outperforms existing methods on event prediction tasks. In the future, we will increase the number of datasets and add more baseline models for comparison. We will also apply our model to different types of event prediction tasks like rumor detection, traffic event prediction, and health event prediction. 
\bibliographystyle{IEEEtran}
\bibliography{main}

\end{document}